\documentclass[twocolumn,showpacs,superscriptaddress,prl]{revtex4}

\usepackage{xspace,psfig,epsfig,amsfonts,amssymb,color}

\begin{document}

\newcommand{\beq}{\begin{equation}} \newcommand{\eeq}{\end{equation}}
\newcommand{\bqa}{\begin{eqnarray}} \newcommand{\eqa}{\end{eqnarray}}
\newcommand{\nn}{\nonumber} \newcommand{\nl}[1]{\nn \\ && {#1}\,}
\newcommand{\erf}[1]{Eq.~(\ref{#1})}
\newcommand{\erfs}[2]{Eqs.~(\ref{#1})--(\ref{#2})}
\newcommand{\crf}[1]{Ref.~\cite{#1}} 
\newcommand{\dg}{^\dagger}
\newcommand{\rt}[1]{\sqrt{#1}\,}
\newcommand{\smallfrac}[2]{\mbox{$\frac{#1}{#2}$}}
\newcommand{\half}{\smallfrac{1}{2}}
\newcommand{\bra}[1]{\langle{#1}|} \newcommand{\ket}[1]{|{#1}\rangle}
\newcommand{\ip}[2]{\langle{#1}|{#2}\rangle}
\newcommand{\sch}{Schr\"odinger} \newcommand{\hei}{Heisenberg} 
\newcommand{\bl}{{\bigl(}} \newcommand{\br}{{\bigr)}}
\newcommand{\ito}{It\^o} \newcommand{\str}{Stratonovich}
\newcommand{\sq}[1]{\left[ {#1} \right]} 
\newcommand{\cu}[1]{\left\{ {#1} \right\}} 
\newcommand{\ro}[1]{\left( {#1} \right)}
\newcommand{\an}[1]{\left\langle{#1}\right\rangle}
\newcommand{\implies}{\Longrightarrow} 
\newcommand{\tr}[1]{{\rm Tr}\sq{ {#1} }} 
\newcommand{\del}{\nabla} \newcommand{\du}{\partial} 
\newcommand{\dbd}[1]{{\partial}/{\partial {#1}}}
\newcommand{\tp}{^{\top}} 
\newcommand{\tbt}[4]{\left( \begin{array}{cc} {#1}& {#2} \\ {#3}&{#4} \end{array}\right)}
\renewcommand{\section}[1]{{\em #1}.---}

\title{Optimal Unravellings for Feedback Control in Linear Quantum
  Systems}

\author{H.\ M.\ Wiseman} \affiliation{Centre for
  Quantum Computer Technology, Centre for Quantum Dynamics, School of Science, 
   Griffith University, Brisbane 4111 Australia}
\author{A.\ C.\ Doherty} \affiliation{School of Physical Sciences,
  University of Queensland, Brisbane 4072 Australia}

\date{24th November, 2004}

\begin{abstract}
  For quantum systems with linear dynamics in phase space much of
  classical feedback control theory applies. However, there are some
  questions that are sensible only for the quantum case, such as:
  given a fixed interaction between the system and the environment
  what is the optimal measurement on the environment for a particular
  control problem? We show that for a  broad class of optimal ({\em state-based}) control
  problems (the stationary {\em Linear-Quadratic-Gaussian} class), 
  this question is a semi-definite program.   
  Moreover, the answer also applies to Markovian ({\em current-based}) feedback. 
\end{abstract}
 \maketitle

Classical feedback control is a large and well established area of
engineering \cite{Jac93,Zhou96}. Quantum feedback control, by
contrast, is still rapidly developing (see Ref.~\cite{WisManWan02} for
a recent review of theory) especially with the notable success of
recent experiments in the field \cite{Smi02,Arm02,Ger04,Hay04}. 
Although problems in quantum control are generally quite
distinct from their classical counterparts, for quantum systems
with linear dynamics in phase space, and with Gaussian noise,
classical optimal feedback control theory can be applied. This idea
was introduced by Belavkin
\cite{Bel87}, and its physical application was 
developed in Refs.~\cite{DohFeedback}. In linear
systems the key differences between the quantum and classical problems
are the limits resulting from measurement backaction noise.   The
significance of this field
is that in many quantum feedback control experiments a linear theory can be used, 
for example in quantum optics \cite{Wis95c,Arm02}, atomic ensembles 
\cite{ThoManWis02,Ger04}, nanomechanics 
\cite{Hop03,Hay04} and cavity QED \cite{Ste04}. 

This Letter contains several important advances in this field. First,
we give a completely general theory of feedback control in linear
quantum systems with a fixed coupling to the environment, allowing for
arbitrary unravellings (ways of monitoring the environment). 
Second, having formulated the conditional evolution under arbitrary unravellings 
as stochastic differential equations for the moments,
we show that there is a {\em stabilizing} solution $W$ for the
 conditioned covariance matrix under weaker conditions than classically.
 Third, we identify which unconditional evolutions allow for such a $W$ to
exist for almost all unravellings, and show that the set of all possible $W$s is set by two 
{\em linear matrix inequalities} (LMIs).
 Fourth, we show that for all stationary problems with a cost function 
 that is quadratic in the system and controller variables, 
 the optimal unravelling for optimal feedback control can be found efficiently using a semi-definite
program. Fifth, we show that if all control constraints are removed, Markovian
feedback control (which is much simpler both conceptually and
experimentally) performs identically to optimal feedback control. We
conclude with an example followed by discussion.

\section{Continuous Markovian Unravellings}
The most general autonomous differential equation for the state of a
quantum system is the Lindblad master equation
\begin{equation}
\hbar \dot{\rho}=-i[\hat H,\rho ]+ \sum_{l=1}^{L}{\cal D}[\hat{c}_l]\rho  \equiv {\cal L}_0 \rho
\label{eq:lindblad}\label{nofbME} %
\end{equation}
Here $\hat H = \hat H\dg$ is the system Hamiltonian, $\cu{\hat c_l}$
is a set of operators that are arbitrary (although for rigour
\cite{Lin76,GorKosSud76} they should be bounded). The action of ${\cal D}[\hat c]$
on an arbitrary operator $\rho$ is defined by 
\beq {\cal D}[\hat
c]\rho \equiv \hat{c}\rho \hat{c}^{\dagger }-\left( \hat{c}^{\dagger
  }\hat{c}\rho +\rho \hat{c}^{\dagger }\hat{c}\right) /2.  \eeq
Lindblad master equations are widely used in atomic, optical, and
nuclear physics \cite{fn1}. 
They can typically be derived if the system is coupled weakly
to an environment that is large (i.e. with dense energy levels).
Because the system typically becomes entangled with the environment,
tracing over the latter typically leads to loss of system purity.
However, there is no necessity to ignore the environment. Indeed,
under the conditions that allow the derivation of a Lindblad master
equation, it is possible to measure the environment continually on a
time scale much shorter than any system time of interest. This {\em
  monitoring} yields information about the system, producing a
stochastic {\em conditioned} system state $\rho_{\rm c}$ that {\em on
  average} reproduces the unconditioned state $\rho$. That is, the
master equation is {\em unravelled} into stochastic quantum
trajectories \cite{Car93b}, with different measurements on the
environment leading to different unravellings.

If we ask that the monitoring also yield an evolution for $\rho_{\rm
  c}$ that is, like that of \erf{eq:lindblad}, continuous and
Markovian, then it must be of the form \cite{WisDio01}
\begin{eqnarray}
\hbar d{\rho}_{\rm c} &=& 
{\cal L}_0 \rho_{\rm c}dt + 
d {\bf z}\dg (t) \Delta_{\rm c} \hat{\bf c} \rho_{\rm c}  +\rho_{\rm c} \Delta_{\rm c}
\hat {\bf c}^{\dagger } d{\bf z}(t) .\label{eq:sme}
\end{eqnarray}
Note that here the $\dagger$ indicates transpose ($\top$) of the
vector and Hermitian adjoint of its components, and $\hat{\bf c} =
(\hat{c}_1,\cdots,\hat{c}_{L})\tp $. We are also using the notation
$\Delta_{\rm c} \hat {o} \equiv \hat o - \an{ \hat o}_{\rm c}$, where
$\an{ \hat o}_{\rm c} \equiv {\rm Tr}[\rho_{\rm c} \hat{o}]$, and we
have introduced $d{\bf z} = (dz_1, \cdots ,dz_{L})\tp$ of
infinitesimal complex Wiener increments \cite{Gar85}. It satisfies
E$[d{\bf z}] = 0$, where E denotes expectation value, and has the
correlations \beq \label{noisepower} d{\bf z} d{\bf z}\dg = \hbar \Theta
 dt , \;\; d{\bf z} d{\bf z}\tp = \hbar \Upsilon dt.  \eeq Here we
have generalized Ref.~\cite{WisDio01} to allow for inefficient
detection by introducing a matrix $\Theta$.
It is convenient to combine $\Upsilon$ and $\Theta$ in an {\em unravelling matrix}
 \begin{equation}
U \equiv  \frac{1}{2}\left( 
\begin{array}{cc}
\Theta+\text{Re}\left[ \Upsilon \right]  & \text{Im}\left[ \Upsilon\right]  \\ 
\text{Im}\left[ \Upsilon\right]  & \Theta-\text{Re}\left[ \Upsilon\right] 
\end{array}
\right),
\end{equation}
The set $\mathfrak{U}$ of valid $U$s is defined by $U\geq 0$, 
$\Upsilon = \Upsilon\tp$, and 
$\Theta={\rm diag}(\theta_1, \cdots ,\theta_L)$, where $\theta_l \in [0,1]$.  This 
can be derived using the method of Ref.~\cite{WisDio01} by splitting each output channel $l$ into 
a portion $\theta_l$ that is observed and a portion $1-\theta_l$ that is unobserved.  
In the case of {\em efficient}
monitoring ($\Theta=I$), \erf{eq:sme} preserves purity and so can be
replaced by a stochastic \sch\ equation \cite{Car93b,WisDio01}.

The measurement results upon which the evolution of $\rho_{\rm c}$ is
conditioned is a vector of complex functions 
\beq \label{defJz} {\bf
  J}\tp dt =\an{\hat{\bf c}\tp \Theta + \hat{\bf c}\dg \Upsilon}_{\rm
  c}dt + d{\bf z}\tp .  \eeq In optics measurements like this (e.g.
homodyne detection) give rise to photocurrents, so we call ${\bf J}$ a
current.

 \section{Linear Systems} \label{sec:LinDyn}
 We are interested in systems of $N$ degrees of freedoms, with the
 $n$th described by the canonically conjugate pair obeying the
 commutation relations $[\hat q_{n},\hat p_{n}] = i\hbar$. Defining a
 vector of operators \beq \hat \mathbf{x}=\left( \hat q_{1},\hat
   p_{1},...,\hat q_{N},\hat p_{N}\right) \tp , \eeq we can write
 $\left[\hat x_{n},\hat x_{m}\right] =i\hbar \Sigma _{nm}$,
 where $\Sigma$ is the $(2N)\times(2N)$  symplectic matrix 
 \beq \Sigma=\bigoplus _{n=1}^{N}  \left(
\begin{array}{cc} 0 & 1 \\ 
-1 & 0 \end{array} \right)  = \Sigma^* = - \Sigma\tp = - \Sigma^{-1}.
\eeq 
 
  For a system with such a phase-space structure it is possible to
  define a Gaussian state. Like its classical
  counterparts, it is determined by its mean vector $\an{ \hat {\bf x}
  }$ and its covariance matrix, which in the quantum case must be
  symmetrized: $ V_{nm}=\left( \langle \Delta \hat x_{n}\Delta \hat
    x_{m}\rangle +\langle \Delta \hat x_{m}\Delta \hat x_{n}\rangle
  \right) /2$.
For such states, a necessary and sufficient condition on the matrix $V$
is the following LMI \cite{Hol75} 
\begin{equation}
V+i\hbar \Sigma /2\geq 0.
\label{eq:uncert} \label{GSLMI}
\end{equation}

In this Letter we are concerned with linear systems; that is, ones for
which $\hat H$ is quadratic, and $\hat{\bf c}$ linear, in $\hat{\bf
  x}$:
\begin{equation}
\hat H=(1/2)\hat \mathbf{x}\tp G\hat \mathbf{x} -\hat \mathbf{x}\tp \Sigma B\mathbf{u}(t)  ,\;\;\hat {\bf c} = \tilde C \hat {\bf x},
\end{equation}
where $G$ is real and symmetric and $B$ is real. The second term in
$\hat H$ is linear in $\hat {\bf x}$ to ensure a linear map between
the time-dependent classical input ${\bf u}(t)$ to the system and the
output current ${\bf J}(t)$.  For such a system, the unconditioned
master equation (\ref{nofbME}) has a Gaussian state as its solution,
with the following moment equations
\begin{eqnarray}
{d\langle \mathbf{\hat{x}}\rangle}/dt  &=&  A\langle \hat \mathbf{x}\rangle +B\mathbf{u}(t) ,  \\
 d{V}/dt &=&AV+VA\tp +D.
\label{eq:dyn}
\end{eqnarray}
Here $A=\Sigma ( G + \text{Im}[ \tilde{C}\dg \tilde{C} ] )$ and $D=
\hbar \Sigma \text{Re}[ \tilde{C}\dg \tilde{C}] \Sigma\tp$. 

 For  {\em conditional} evolution of linear quantum systems 
it is convenient to recast the complex current ${\bf J}$ of \erf{defJz} as a real current with
{\em uncorrelated} noises: 
\beq {\bf y} \equiv (\hbar U)^{-1/2} \left(
\begin{array}{c}
\text{Re}\left[ {\bf J} \right]  \\ 
\text{Im}\left[ {\bf J} \right] 
\end{array}
\right) = C\langle \mathbf{\hat x}\rangle +\frac{d\mathbf{w}}{dt}.
\end{equation}
Here $ C = 2 (U/\hbar)^{1/2} \bar{C}$, where $\bar{C}\tp \equiv 
\left( \text{Re}[ \tilde{C}\tp] , \text{Im}[ \tilde{C}\tp] \right)$, while 
$d\mathbf{w}$ is a vector of real Wiener increments satisfying $d\mathbf{w}d\mathbf{w}\tp =Idt$ \cite{Gar85}. The state $\rho_{\rm c}$
conditioned on $\mathbf{y}(t)$ is found from  \erf{eq:sme}. For linear systems   
this conditioned state is Gaussian, and is analogous to the \emph{a
  posteriori} probability distribution propagated by the classical Kalman filter
equations, with $d\mathbf{w}$ as the 
innovation \cite{Jac93}. 

 Taking expectation values and using the  \ito\ calculus \cite{Gar85}, \erf{eq:sme} yields the  conditional moment equations
\begin{eqnarray}
d\langle \hat \mathbf{x}\rangle_{\rm c}=\left[ A \langle \hat \mathbf{x}\rangle_{\rm c} +B\mathbf{u(%
}t\mathbf{)}\right] dt 
+ \left( V_{\rm c}C\tp +  \Gamma\tp \right) d\mathbf{w}  \label{eq:kfilt1}\\
\dot{V}_{\rm c} =AV_{\rm c}+V_{\rm c}A\tp +D 
-(V_{\rm c}C\tp + \Gamma\tp )(CV_{\rm c}+ \Gamma) , \label{eq:kfilt2} \label{MRE}
\end{eqnarray}
where $\Gamma = - (\hbar U)^{1/2}S\bar{C}\Sigma\tp$, where 
\beq S= {\tbt{0}{I}{-I}{0}}.
\eeq
The stochastic term in $d\langle \mathbf{x}\rangle_{\rm c}$ and the final term in $\dot{V}_{\rm c}$ describe the conditioning on the measured current ${\bf y}$. 

Note that the equation for $V_{\rm c}$ is deterministic and independent
of the measurement results.
The final term causes a 
reduction in uncertainty about the system state (that is, a reduction in the eigenvalues of $V_{\rm c}$). 
If \erf{MRE} has a {\em stabilizing solution} \cite{Zhou96}, then all initial conditions will asymptote to the same steady state. We will notate a stabilizing solution $V_{\rm c}^{\rm ss}$ as $W_U$ to emphasize that it depends upon the unraveling $U$. Considering efficient detection for simplicity, $W_U$ is the solution of  
\beq \label{AMRE}
0 =\Omega  W_U+W_U \Omega \tp -W_UC\tp CW_U  + E E\tp.
\eeq
Here $\Omega  = \Sigma[ G + \bar{C}\tp S(2U-I)\bar{C}]$ is Hamiltonian drift, while
$E=\hbar \Sigma C\tp/2$, manifesting the measurement back-action
noise resulting from having ${\bf y} \propto C\hat{\bf x}$. For a general classical problem, $\Omega$, $C$, and $E$ would be unrelated.

A {\em necessary} condition for \erf{AMRE} to have a stabilizing solution is for $(C,\Omega )$  to be {\em detectable} \cite{Zhou96}, which means  
\beq \label{detectability}
C{\bf v} \neq 0\; \forall\, {\bf v}:\, \Omega {\bf v} = \lambda{\bf v} \textrm{ with Re}(\lambda) \geq 0.
 \eeq
That is, all the degrees of freedom that are not strictly stable 
under the drift matrix $\Omega$ contribute to
the signal $C{\bf x}$. Classically this becomes a {\em sufficient} condition if we have also  
$(E\tp,\Omega \tp)$  detectable \cite{Zhou96}. 
Quantally, this extra assumption is unnecessary because of back-action. To see this, note that from \erf{detectability}, 
$(C,\Omega )$ detectable  $\implies (C\Sigma\tp,\Omega \Sigma\tp)$ detectable by  the invertibility of $\Sigma$; 
then by the symmetry of $\Omega\Sigma\tp$ this implies $(C\Sigma\tp,\Sigma\Omega\tp)$ detectable, which implies $(E\tp,\Omega \tp)$ detectable as desired. That is, unlike the classical case, detectability of $(C,A)$ [which is equivalent to detectability of $(C,\Omega)$]  is sufficient for $W_U$ to be a stabilizing solution of \erf{eq:kfilt2}.
 
 \section{Possible Conditional States} 
Let the $U=I/2$ unravelling be detectable [i.e. let $(\bar{C},A)$ be detectable]. Then the 
$W_U$s that result from detectable unravellings can be shown to be dense in the set $\mathfrak{W}=\cu{W_U:U\in\mathfrak{U},W_U<\infty}$ of 
 finite  solutions $W_U$ to \erf{AMRE}. 
 Moreover if the $U=I/2$ unravelling is not detectable then no unravelling is, because $\bar{C}{\bf v}={\bf 0} \implies C{\bf v}={\bf 0}$. Thus, for all practical purposes, any $W_U \in \mathfrak{W}$ will be a  stabilizing solution if and only if $(\bar{C},A)$ is detectable (an unravelling-independent condition). 

The set $\mathfrak{W}$ can be determined as follows. Note that from the positivity of the final term in \erf{eq:kfilt2}, 
\beq \label{LMI:cond}
D+A W_U + W_U A\tp  \geq 0
\eeq
is a necessary condition on $W_U$. This condition means that a state with $V_{\rm c}=W_U$ will evolve unconditionally to a state with $V = W_U + dt(D+A W_U + W_U A\tp) \geq W_U$, in other words to a convex (Gaussian) combination of states with $V=W_U$.  By obtaining information from the bath, the system can thus be steered back to a state with $V_{\rm c}=W_U$ \cite{fnWisVac01}.  That is, \erf{LMI:cond} is also a sufficient condition on $W_U$ for the existence of some $U$ 
which will yield $V_{\rm c}^{\rm ss} = W_U$, provided that $W_U$ is a valid covariance matrix ($W_U\geq 0$ classically, and 
$W_U+i\hbar \Sigma /2\geq 0$ quantally).
 Moreover,  given a $W_U$ that satisfies the LMIs (\ref{LMI:cond}) and (\ref{GSLMI}) (with $V\to W_U$), a (not necessarily unique) unravelling $U$ that will generate it can be found by solving 
\beq \label{LME}
\hbar R\tp U R = D+A W_U + W_U A\tp  ,
\eeq
where $R = 2\bar{C}W_U/\hbar + S\bar{C}\Sigma$. This comes from substituting the expressions for $\Omega$, $C$ and $E$ into \erf{AMRE}.


\section{Optimal Quantum Control}
In feedback control,  ${\bf u}(t)$
depends on the history of the measurement record ${\bf y}(s)$ for $s<t$.   
The typical aim of control over some interval $[t_0,t_1]$ is to minimize the expected value of a {\em cost function} \cite{Jac93}, the integral of the sum of positive functions of ${\bf x}(t)$ and ${\bf u}(t)$ for $t_0<t<t_1$. For any such problem the {\em separation principle} holds: the optimal control ${\bf u}(t)$ depends only upon 
the observer's state of knowledge about the system. That is, in the quantum case, the measurement record ${\bf y}(s)$ for $t<s$ is irrelevant except in how it determines $\rho_{\rm c}(t)$. A special case of interest is that of Linear-Quadratic-Gaussian (LQG) control \cite{Jac93}: a {\em linear} system with a {\em quadratic} cost function and having {\em Gaussian} noise. LQG control has the additional property of {\em certainty equivalence}:  the only property of $\rho_{\rm c}$ required is $\an{\hat {\bf x}}_{\rm c}$. 
Moreover, the optimal ${\bf u}$ is linear in this mean: 
\beq
{\bf u}(t) = -K(t) \an{\hat{\bf x}}_{\rm c}(t).
\eeq

In this Letter we specialize to the case of time-independent cost functions, and $(t_1 - t_0) \to \infty$.  That is, we wish to minimize $m = {\rm E}[h]$ in steady state, where
\beq
h = \an{ \hat{\bf x}\tp P \hat{\bf x} }_{\rm c} + {\bf u}\tp Q {\bf u},
\eeq
where  $P\geq 0$ and $Q\geq 0$.  Note that in steady state 
\beq \label{usefulforMarkov}
{\rm E}_{\rm ss}[\an{ \hat{\bf x}\tp P \hat{\bf x} }_{\rm c} ] = {\rm tr}[W_U P] + 
{\rm E}_{\rm ss}[\an{\hat{\bf x}}_{\rm c}\tp P\an{\hat{\bf x}}_{\rm c}],
\eeq
just as in classical control. Assuming that $W_U$ is stabilizing, $Q^{-1}B$ exists, and $(B\tp,A\tp)$ and $(P,A\tp)$ are detectable, there is a stable optimal control law: $K=Q^{-1}B\tp Y$ \cite{Zhou96}. Here $Y$  is independent of $U$, satisfying 
$P + A\tp Y + Y A = YBQ^{-1}B\tp Y$. 
The resulting (minimum) cost is \cite{Jac93}
\beq \label{costlinearinW}
m_{\rm opt} = {\rm tr} [YBQ^{-1}B\tp YW_U] + {\rm tr} [Y D].
\eeq

 \label{sec:SDP}

The significance of \erf{costlinearinW} is that it makes it explicit that 
the choice of unravelling $U$ (which determines $W_U$) affects the cost of the control. 
Thus we can ask the following: given an open system with 
dynamics described by the drift $A$ and diffusion $D$ matrices, 
what is the optimal way to monitor the bath?
Note that classically 
this would be a nonsensical question as the unconditioned evolution described
by $A$ and $D$ would 
not proscribe the measurements that can be made on the system in any
way. But for quantum systems $W_U$  will be positive definite because 
there is no measurement without disturbance, and so $m_{\rm opt} \neq 0$ 
even for zero control cost ($Q \to 0$). 

Finding the optimum unravelling ($U$) is computationally efficient in the system size $N$. This is because at its heart is the semi-definite program \cite{VanBoy96} of minimizing a linear function of $W_U$ (\ref{costlinearinW}) subject to the constraints of the LMIs (\ref{GSLMI}) (with $V\to W_U$) and (\ref{LMI:cond}). The optimum unravelling $U$ is then found from \erf{LME}. 

\section{Markovian Quantum Control} If we remove all constraints on the control by making $B$ full rank  and letting 
$Q\to 0$, \erf{costlinearinW} simplifies to
\beq \label{tracePV}
m_{\rm opt} = {\rm tr} [P W_U] 
\eeq
It turns out that the solution to this problem (the optimal $U$) is relevant not only for optimal control, but also for Markovian control as introduced by Wiseman and Milburn \cite{WisMilFeedback}. Conceptually and experimentally this is a much simpler form of feedback, as it entails making the time-dependent Hamiltonian linear in the instantaneous output ${\bf y}(t)$ [rather than the Kalman-filtered output, $\an{\bf x}_{\rm c}(t)$]. That is, in the context of the linear system, 
\beq
\mathbf{u}(t)  = F(t) \mathbf{y}(t)
\eeq
Note that ${\bf y}(t)$ has unbounded variation, so doing Markovian control is no less onerous than doing optimal control with unbounded $K(t)$ as occurs for $Q\to 0$.   With $B$ invertible we can choose 
$BF = - W_UC\tp  - \Gamma\tp$. This makes 
\erf{eq:kfilt1} deterministic in the limit $t\to\infty$:
\beq \label{MarkovStable}
d\langle \hat \mathbf{x}\rangle_{\rm c} = M \langle \hat \mathbf{x}\rangle_{\rm c}dt,
\eeq
where $
M \equiv A - W_U C\tp C - \Gamma\tp  C$. 
This is the generalization of the optimized Markovian feedback strategy identified for 1-dimensional systems in Refs.~\cite{WisMilFeedback}.
As long as $W_U$ is stabilizing,  $M$ will be strictly stable \cite{Zhou96} so that  the solution of \erf{MarkovStable} will asymptote to ${\bf 0}$. Thus from \erf{usefulforMarkov} the cost will again be given by \erf{tracePV}, and the optimal unravelling found as above.

\section{Example} Consider a system with $N=L=1$ described by the master equation 
\beq
\hbar \dot{\rho} = -i[(\hat q\hat p+\hat p\hat q)/2,\rho] + {\cal D}[\hat q+i\hat p] \rho,
\eeq
where the output arising from the second term may be monitored. This could be realized in quantum optics as a damped cavity (harmonic oscillator in the rotating frame) containing an on-threshold parametric down converter \cite{Car93b}, with $p$ the squeezed quadrature.

In this case we have $\tilde{C} = (1, i)$, so the drift and diffusion matrices evaluate to
$A = {\rm diag}(0,-2)$, $D=\hbar I$. 
Writing the conditional steady-state covariance matrix as
\beq
W_U = \frac{\hbar}{2} {\tbt{\alpha}{\beta}{\beta}{\gamma}},
\eeq
the LMIs (\ref{GSLMI})  and (\ref{LMI:cond})  become
\beq
\tbt{\alpha }{\beta+i}{\beta -i} {\gamma} \geq 0,\;\;
\tbt{1}{-\beta} {-\beta}{1-2\gamma} \geq 0 .
\eeq

Now say the aim of the feedback control is to produce a stationary state where $q = p$ as nearly as possible. A suitable cost function to be minimized is
$\an{(\hat q - \hat p)^2}_{\rm ss}$. 
That is, ignoring any control costs, we have 
\beq P = {\tbt{1}{-1}{-1}{1}}, Q \to 0.
\eeq
 In the optics case it is simple to displace the system in its phase space by application of a coherent driving field \cite{Car93b}. That is, we are justified in taking $B$ to be full rank. Furthermore,  any quadratic cost function will be minimized for a pure state so we may assume that $\alpha\gamma = 1 + \beta^2$.  Thus the $m$ achievable by optimal or Markovian control is 
$m = {\rm tr}[P W_U] =  [\ro{1+\beta^2}/{2\gamma} + \gamma/2-
\beta] \hbar$, 
constrained only by $0< \gamma \leq (1-\beta^2)/2$. The minimum is found numerically to be 
$m \approx 1.12 \hbar$ at $\beta \approx 0.248$ and $\gamma = (1-\beta^2)/2$. 
Proceeding as described above, we find $\sqrt{\hbar}C/2 = -\Gamma/\sqrt{\hbar} = U^{1/2}$, with the optimal unraveling 
\beq
U  = \tbt{\cos^2\phi}{\cos\phi\sin\phi}{\cos\phi\sin\phi}{\sin^2\phi} \textrm{ for }\phi \approx 0.278\pi.  
\eeq 

Optically, this unravelling corresponds to homodyne detection with $\phi$ being the local oscillator phase.
For Markovian feedback, this gives gives the drift matrix
\beq
M = A - 4W_UU/\hbar - 2U  \approx \tbt{-2.94}{-3.50}{-1.65}{-3.97},
\eeq
which is strictly stable as required.

To conclude, we have shown that even for  quantum systems that are linear (and so have a classical analogue) the constraints of quantum theory affect the basic structure of feedback control problems. In particular, we have formulated a natural question --- the optimal unravelling for stationary LQG control problems --- that has no classical analogue. Moreover, these constraints also lead to an efficient algorithm to answer this question.  This theory applies to any linearizable system with quantum-limited monitoring. No doubt further fundamental aspects of  control for such systems still await discovery.  

We thank the ARC and the State of Queensland for support, and K. Jacobs and H. Mabuchi for discussions.


\begin{thebibliography}{99}
\bibitem{fn1}
 Non-Lindblad-form master equations (including integro-differential
equations and non-autonomous differential equations) are also widely used,
when the Markov approximation is invalid. For such systems the unravelling of the master equation into stochastic trajectories cannot be interpreted using standard quantum measurement theory 
 --- see Ref.~\cite{GamWis03}. The 
theory of this Letter is thus limited to master equations of the form of \erf{nofbME}.
\end{thebibliography}
\end{document}